\documentclass[xcolor=pdftex,dvipsnames,usenames,prd,nofootinbib,12point]{revtex4}
\usepackage{amsmath}
\usepackage{slashed} 

\input{texdefinitions}

\newcommand{\bfr}{{\bf r}}

\newcommand{\ben}{\begin{displaymath}}
\newcommand{\een}{\end{displaymath}}

\newcommand{\eq}[1]{Eq.~(\ref{#1})}

    \newcommand {\boldsigma}{\mbox{\boldmath$\sigma$}}

\begin{document}

\title{ Non-Perturbative Lepton  Sea Fermions in the Nucleon and the Proton Radius Puzzle\\\hskip7cm NT@UW-14-27}
\author{Gerald A. Miller}
\address{University of Washington, Seattle WA 9819-1560}
\date{\today}
\begin{abstract} 
A potential explanation [U.~D.~Jentschura,
  Phys.\ Rev.\ A {\bf 88}, 062514 (2013)] of the proton radius puzzle originating from  the non-perturbative lepton-pair content of the proton is studied. Well-defined  quantities that depend on this 
lepton-pair content are evaluated. Each   is found to be of the order of $10({\alpha\over \pi})^2$, so that we find such a lepton-pair content exists in the proton. However, we argue 
that this  relatively large result  and general features of loop diagrams   rule out 
the possibility of  lepton-pair content as an explanation of the proton radius puzzle. The contributions of   a class of potential explanations of the proton radius puzzle  (for which the dependence on
the $\mu p$ relative distance is as contact interaction)  are  shown to be  increase very rapidly with atomic number.
\end{abstract}
\maketitle
\vspace{-3.0em}

\section{Introduction}

Recent high precision experimental studies of muonic hydrogen
 \cite{PoEtAl2010,AnEtAl2013}
obtain a value of the proton radius that  is about 4\%  smaller than that obtained from ordinary electronic hydrogen. The problem of understanding this difference has become known as the proton radius puzzle, and has generated a vast array of possible solutions, see the review \cite{Pohl:2013yb}. 
One of the novel suggested solutions~\cite{Jentschura:2014ila} is that a non-perturbative feature of the proton's structure, namely  the
possible presence of light sea fermions
as constituent components of the proton, could account for the difference in the extracted radii. In particular, it is argued that the assumption that   the presence  of $2.1  \times 10^{-7}$ light sea positrons per  
quark,  leads to an
  an extra
term in the electron-proton versus muon-proton interaction, which has the right
sign and magnitude to explain the proton radius puzzle .  

\begin{figure}[h]
\includegraphics[width=5cm,height=5cm]{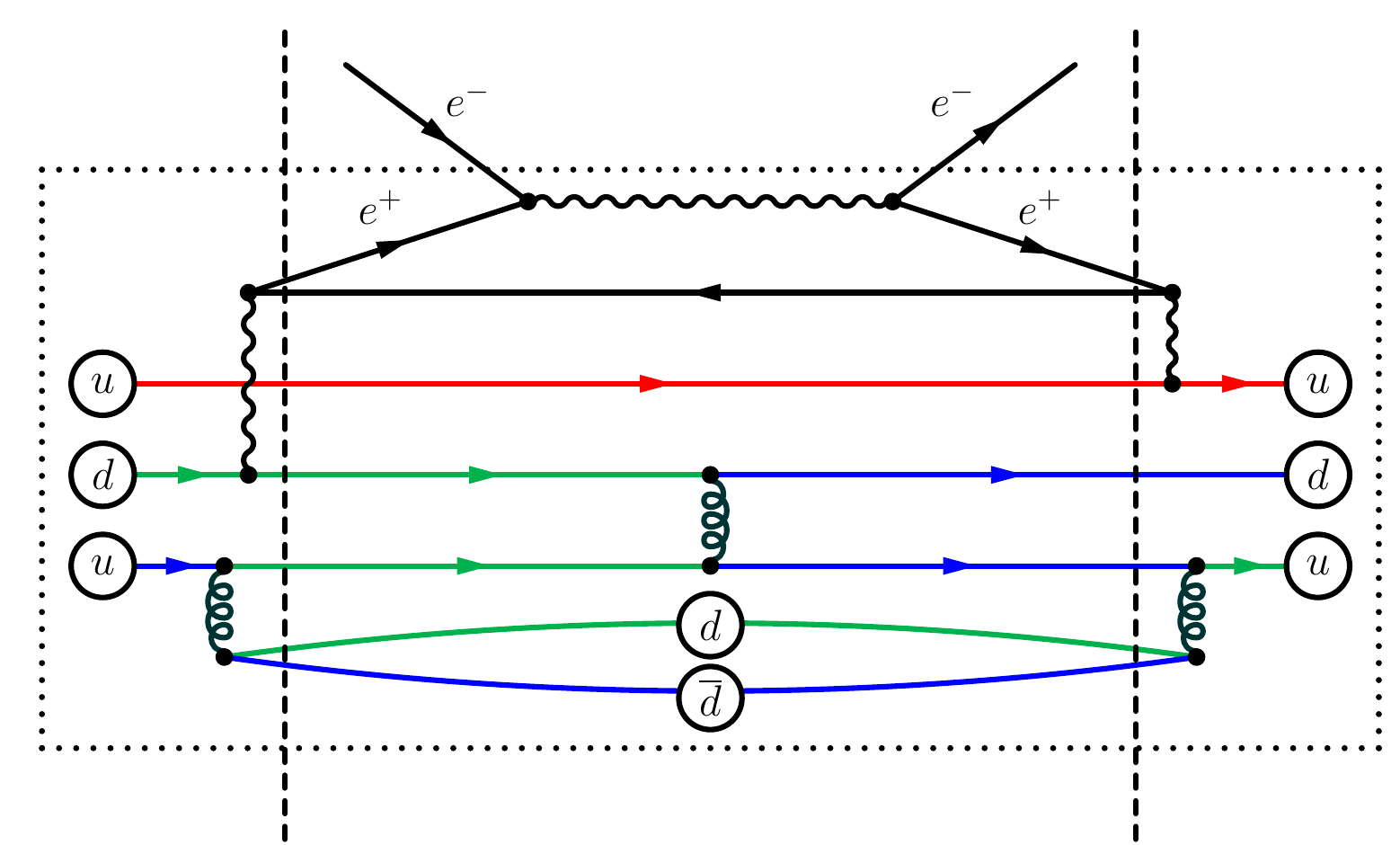}
\caption{(color online)   Typical Feynman diagram of Ref.~\cite{Jentschura:2014ila} illustrating the virtual
annihilation of a bound electron with a ``light sea lepton'' (positron) inside
the proton.  The up ($u$) and down ($d$) quarks, which carry non-integer charge
numbers, interact electromagnetically.
The vertical dashed lines indicate  the lepton-sea component of the proton wave function. 
The  light sea lepton
that annihilates with the bound electron.
An example of a non-perturbative  quantum chromodynamic
(QCD) interaction via a blue-antigreen gluon also is
indicated in the figure.   }\label{f1fig}\end{figure}

The basic idea is that a bound electron may annihilate with a positron that is part of the non-perturbative $e^-e^+$ cloud of the proton. The annihilation leads to a virtual photon, which in turn decays to a bound electron 
and a positron that is also part of the  $e^-e^+$ cloud of the proton.  See Fig.~\ref{f1fig}.  The term non-perturbative here refers to a component  of the proton Fock-space wave function that can be seen  at small values of   momentum transfer. If one could take a snapshot of a proton in isolation one would see electron-positron or muon-anti-muon pairs pop in and out of existence. 
This effect is therefore different from the generation of pairs by evolution in momentum transfer that is akin to the source of the $q\bar{q}$ sea of perturbative QCD. 
 
 A natural question to ask is whether or not this diagram is part of a contribution that is already included. In particular,  a look at Fig~\ref{f1fig}  might lead one to conclude that the intermediate baryonic state is that  of
 an excited nucleon.   If so,  this  diagram would be  a particular time-ordering of the proton polarizability contribution to the two-photon exchange diagram.
   However, if the lepton pair is a specific Fock-space  component of  the complete  proton wave function (including QED effects)   one may  argue~\cite{Jentschura:2014ila} that the intermediate state is part of the proton wave function, and therefore not part of the proton polarizability contribution.  Here we accept this argument and seek to determine its logical consequences.

\begin{figure}[h]
\includegraphics[width=6cm,height=6cm]{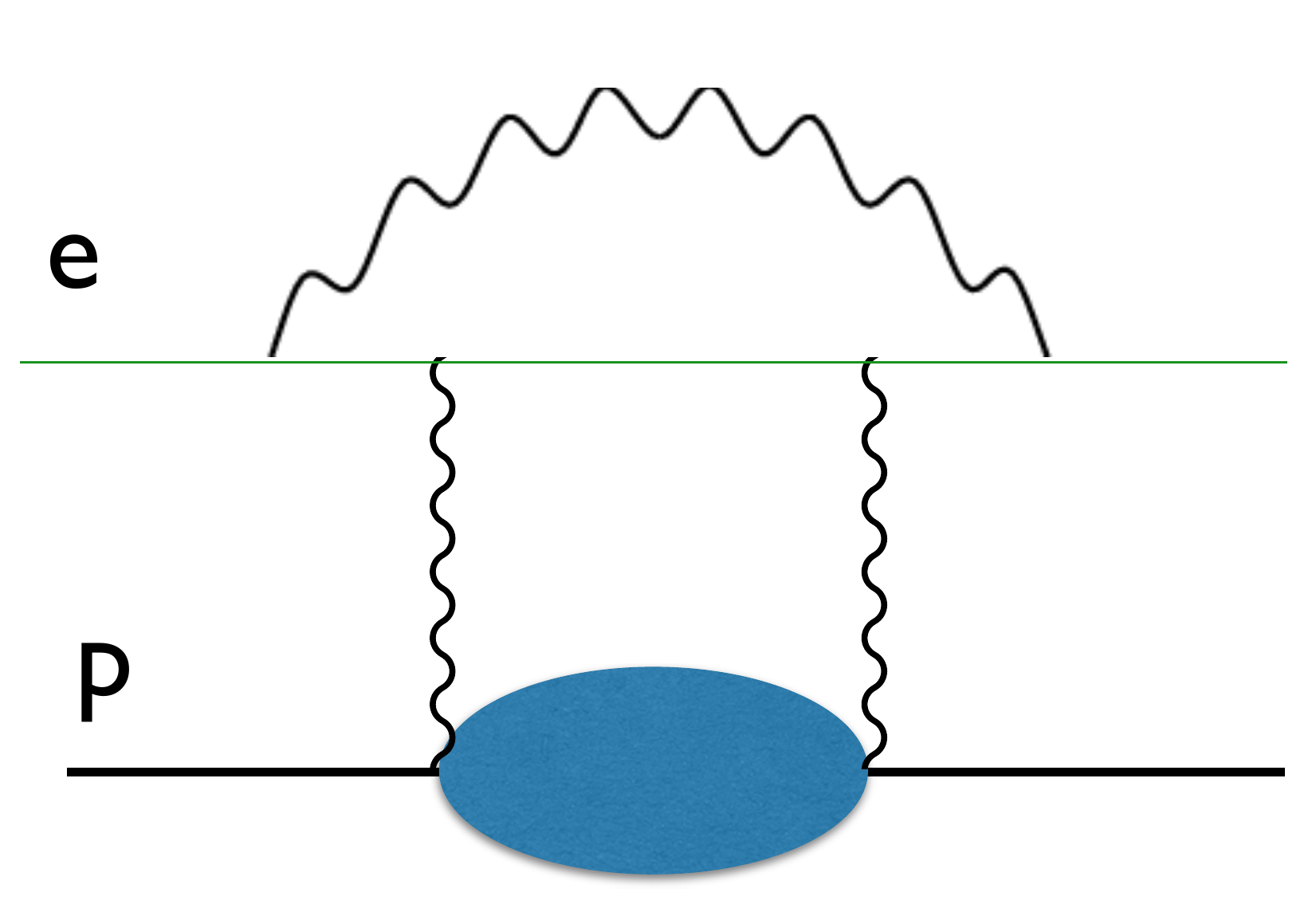}
\caption{(color online)  Radiative correction to two photon exchange. 
 Including the full gauge invariant set, including crossing the photons arising from the proton (p) and all of the one-loop self-energy terms on the electron(e), is implied.}\label{radcorr}\end{figure}

This acceptance comes with severe   difficulties. If one adopts this Fock-space approach, one must recognize that computing the contribution of a given component may only   correspond to computing a  particular
time ordering of a Feynman diagram, with the consequent obligation to find any  remaining time ordering terms of the same order in $\alpha$. To be specific, the term of Fig.~1 corresponds to a particular time ordering of the graphs that give the radiative corrections to the general two photon exchange diagrams. See for example, Fig.~\ref{radcorr}.
In this figure, the intermediate blob is meant to contain the proton and all of its excited states. Thus this graph contains the time-ordering that is included in 
Fig.~1. One may immediately estimate the size of the contribution of this term  to the energy to be of order $\alpha/\pi$ times the polarizability correction, and is therefore negligible.  The explicit calculations for muonic hydrogen are in accord with this estimate~\cite{eides,Martynenko:2001qf}. 
For electronic hydrogen the contribution is again of order $\alpha/\pi$ times the polarizability correction, which is relatively much less important for electronic hydrogen than for
muonic hydrogen\cite{eides}. 
 This means that a correct evaluation of the  contribution of the effects of the proton's lepton-pair content must give a negligible result.

 We pursue the idea of Ref~\cite{Jentschura:2014ila}  even though the result of any correct calculation will be that the effect is negligible. This is because investigating the lepton-pair content of the nucleon is interesting in its own right  and because it useful to see how the correct result arises from the
Fock-space idea. The argument of Ref~\cite{Jentschura:2014ila} starts by considering the photon annihilation term occurring in 
 positronium. This term   leads to the effective interaction~\cite{BeLiPi1982vol4}:
\begin{equation}
\delta H = \frac{\pi \alpha}{2 m_e^2} \, 
\left( 3 +  \boldsigma_+ \cdot \boldsigma_-
\right)  \, \delta(\bfr) \,.\label{one}
\end{equation}
This Hamiltonian gives
a nonzero interaction of the bound electron and the light sea positron
if their spins add up to one.  Ref.~\cite{Jentschura:2014ila} assumes  that the
electron-positron pairs within the proton are not polarized  and replaces
$\vec \sigma_+ \cdot \vec \sigma_- \to 0$ after averaging over the polarizations
of the sea  leptons. Further it is argued~\cite{Jentschura:2014ila} that for atomic (electronic) hydrogen, the additional
interaction of the electron with the proton due to the annihilation channel 
takes the form
\begin{equation}
H_{\rm ann} = \epsilon_p \, \frac{3 \pi \alpha}{2 m_e^2} \, \delta(\bfr) \,,\label{ann}
\end{equation}
where $\epsilon_p$ measures the amount of electron-positron pairs
within the proton. For muonic hydrogen, the effect is expected
to vanish because the dominant contribution to the sea leptons
comes from the lightest leptons, namely, electron-positron pairs
and thus the annihilation channel is not available. Ref.~\cite{Jentschura:2014ila} finds that a value of 
$\epsilon_p = 2.1 \times 10^{-7} $ is sufficient to account for the different values of the  extracted radii. Note that the $1/m_e^2$ dependence occurs as the result of assuming that both the electron and positron are at rest.   

The hypothesis of sea leptons and the use of \eq{ann} raises a number of interesting questions. 
Is the  quantity $\epsilon_p$ well defined? If so, what is its likely range of values? Any positron in the non-perturbative nucleon  sea is not likely to be at rest, so one could also ask if the \eq{ann} is applicable at all. Does the term of \eq{ann}  really exist?
Our purpose here is to investigate these questions.

Here is an outline of the remainder of this paper. The difficulties in computing and defining $\epsilon_p$ are discussed in Sect.~II. Observables that depend on lepton-pair content 
 and can be computed in a gauge invariant manner are discussed in Sects.III and IV. The results of these two sections are analyzed and used to understand more detailed loop calculations 
Sec.~V. The nuclear dependence of all  models in which the contribution to the Lamb shift arises from an interaction containing a Dirac delta function in the lepton-nucleon separation is studied in 
Sec. ~VI.  Some concluding remarks are made in Sect. VII.
\section{Attempting to count positrons in the proton}

We might proceed to count positrons  by defining an  anti-lepton number current density operator, $J^\mu_{\bar{l}}$,
\bea
 J^\mu_{\bar{l}}=\bar{\psi}_{\bar{l}}\gamma^\mu\psi_{\bar{l}},\label{pcount}\eea
and take its expectation value in the physical proton wave function. The operator  $J^\mu_{\bar{l}}$ is defined to act only on anti-leptons ($\bar{l}$), so its 0'th component is the anti-lepton density.
We may take the expectation value of this operator in the proton by assuming that the lepton pair arises from the interactions on a single quark, as in
Fig.~2.  In that diagram $j$ represents  the incoming  momentum, $p$ and $p'=p+j$  are respectively the incoming and outgoing quark momenta, $m$ and $m_q$ are respectively the lepton mass and the 
constituent quark mass, $q$ is the momentum flowing into the inner loop, and  $k$ is the momentum on one branch of the inner loop. 

It is necessary to discuss why we focus on a single quark. In general the momentum in the loop $q$ is very large. Any  operator that removes   a  large  momentum from one quark 
and gives a large momentum to another quark, suffers a vastly  reduces matrix element in the proton wave function because each quark can only support a momentum of the 
order of the inverse radius of the proton, or about 200 MeV/c. 
There are significant implications to this single-quark dominance see below in Sect. \ref{nuc}.

  \begin{figure}[h]
\includegraphics[width=7cm,height=6cm]{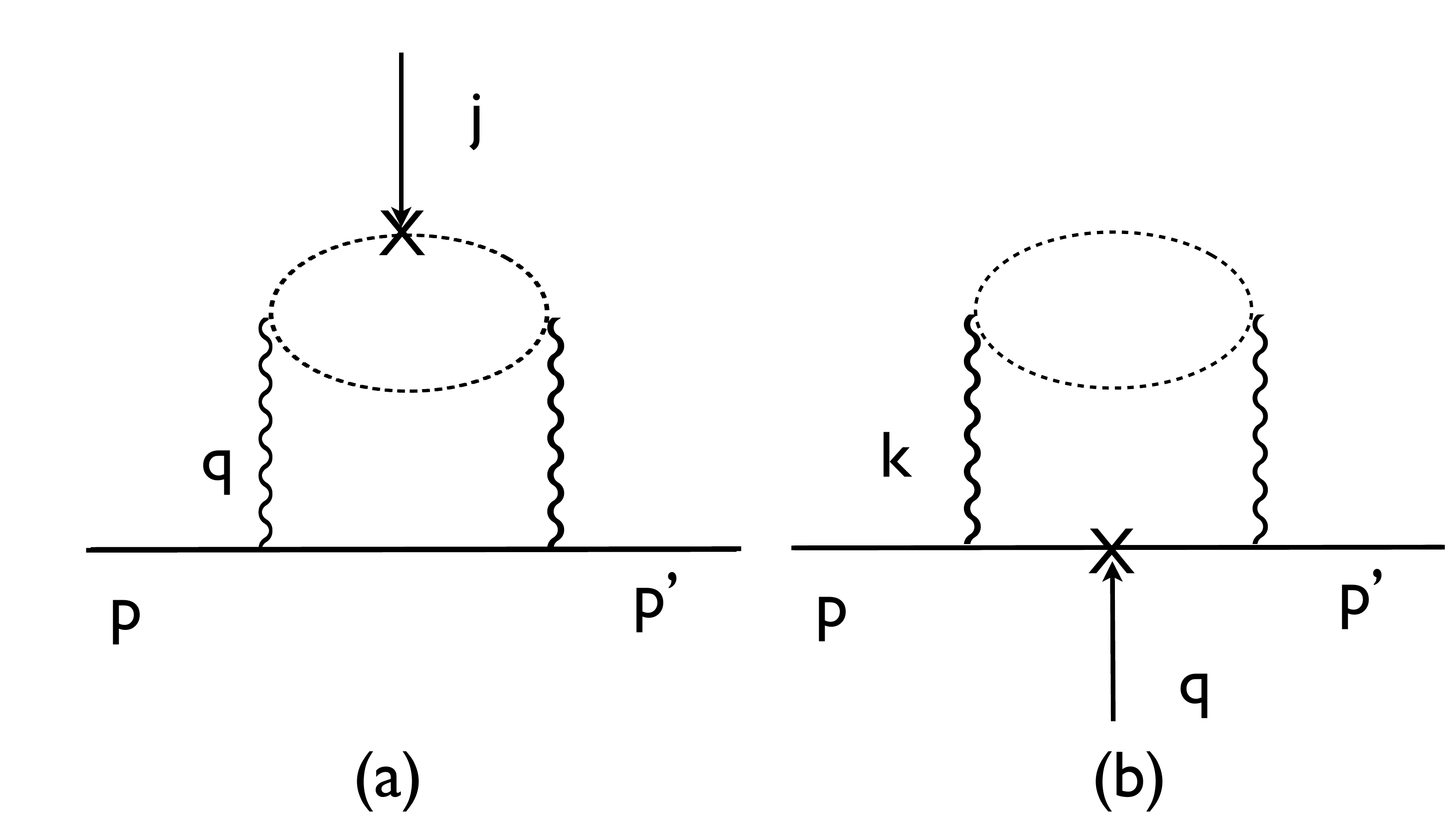}
\caption{(color online) The solid line represents a single quark. (a)  Diagram  for counting anti-leptons. The notation X with the arrow coming into it represents the operator $J^\mu$ bringing in a four-momentum $j$.
(b) Lepton-pair contribution to the vertex function. }
\label{ours}\end{figure}

One immediate problem with the amplitude of Fig.~\ref{ours}  is that the integrations over the virtual lepton momenta contain  an ultraviolet   divergence. This means that an unknown 
 counter term is needed to determine a value, and predictive power is lost.
Furthermore, the use of of \eq{pcount} 
does not respect current conservation, because $J^\mu_{\bar l}$  does not act on all of the charges. Thus  any result would be gauge-dependent. These two 
drawbacks lead us to conclude that 
 the short answer to the question, ``is the quantity $\epsilon_p$ well defined?",
is simply {\it no}.  Thus the validity of  \eq{ann}, which contains a factor of $\epsilon_p$, is questionable. 

However, one may compute matrix elements that depend on the pair content  that are free of ultraviolet  divergences and are gauge invariant. This is the 
procedure  adopted here. 


\section{Anomalous Magnetic Moment}

The goal of this section is to find a simple gauge invariant quantity free of ultraviolet and infrared divergent terms,  that depends on the lepton-pair content of the proton and to evaluate that quantity. Such an example of a 
 leading-order effect of pairs is shown  in Fig.~3b, in which the $X$ represents the usual electromagnetic current operator.
 The intermediate electron line exists only if the lepton-pair exists, and computing only the contribution to the magnetic moment  insures that  gauge invariance is maintained without encountering an ultraviolet or infrared divergence.
 The effects of this diagram have been evaluated long ago, so our purpose is merely   to find an illustrative example.
 
 The effect of pairs 
 is accounted for by dressing the photon propagator using~\cite{PS}
 \bea
{-i g_{\mu\nu}\over q^2}\rightarrow {-i g_{\mu\nu}\over q^2}{1\over 1-\hat{\Pi}_2(q^2)}\approx  {-i g_{\mu\nu}\over q^2}(1+\hat{\Pi}_2(q^2)),\label{pi}
\eea
for a virtual photon of four-momemtum $q$, with 
\bea 
\hat{{\Pi}}_2(q^2)=
{-2\alpha\over\pi} \int_0^1dz\,z(1-z)\log{m_l^2\over m_l^2-z(1-z)q^2},\label{trick}
\eea
where $m_l$ is the lepton mass and the subscript 2 stands for second order in $e$.
It is useful to  rewrite this term  as
\bea 
\hat{\Pi}_2(q^2)=
{-2\alpha\over\pi}\int_0^1dz\,z(1-z)\, q^2\int_0^1{d\lambda\over\lambda} \,{1 \over \hat{m}_l^2-q^2},\label{good}
\eea
where
 \bea \hat{m}_l^2\equiv{m_l^2\over z(1-z)\lambda}.\eea
The use of \eq{good} allows the evaluation  of a Feynman diagram involving $\Pi_2$ via the usual technique  of combining denominators. 

The order $e^4$ correction term to  the  vertex function, $\delta\Gamma^\mu(q)$ is given by using \eq{pi} in the standard expression~\cite{PS} for the $e^2$ term:
\bea \delta \Gamma^\mu(q)=2ie^2\int {d^4k\over (2\pi)^4}{\bar{u}(p')\left (\slashed{k}\gamma^\mu\slashed{k}'+m^2\gamma^\mu-2m(k+k')^\mu\right)u(p)\over \left((p-k)^2+i\epsilon)({k'}^2-m^2+i\epsilon)(k^2-m^2+i\epsilon)\right)}\hat{{\Pi}}_2\left((p-k)^2\right),\label{dg}
\eea
where $p$ is the initial quark momentum, $p'=p+q,\,k'=k+q$ and $m$ is the quark mass, which we take at a constituent  value of one-third of the mass of a proton. The use of \eq{good} in \eq{dg} gives
the result:
\bea
 \delta \Gamma^\mu(q)=2ie^2{2\alpha\over\pi}\int_0^1dz\,z(1-z)\int_0^1{d\lambda\over \lambda}\int {d^4k\over (2\pi)^4}{\bar{u}(p')\left (\slashed{k}\gamma^\mu\slashed{k}'+m^2\gamma^\mu-2m(k+k')^\mu\right)u(p)\over \left((p-k)^2- \hat{m}_l^2+i\epsilon)({k'}^2-m^2+i\epsilon)(k^2-m^2+i\epsilon)\right)}.
 \eea

 The above correction term  contributes to  both the Dirac and Pauli form factors of the quark. Evaluating the Dirac form factor requires
 treatments of infrared and ultraviolet divergences that are not present in the Pauli form factor. We therefore compute only the contribution to the 
 Pauli form factor. Then we proceed by  combining denominators, and integrating over the four-momentum and two of the Feynman parameters and $\lambda$. 
 
 The relevant momentum transfer for atomic physics is the inverse of the Bohr radius, this is much, much  smaller than  the inverse of the quark mass. Therefore
 we need only evaluate $ \delta \Gamma^\mu(q)$ at $q^2=0$.
 We find the following result 
 \bea&
 \delta \Gamma^\mu(q^2=0)= 2({\alpha\over \pi})^2\,\bar{u}(p'){i\sigma^{\mu\nu}q_\nu\over 2m}u(p)I(m^2/m_l^2)\\
 &I(m^2/m_l^2)=\int_0^1dz\,z(1-z)\int_0^1dz'\,\log\left(1+{(1-z')^2z(1-z)\over z'}{m^2\over m_l^2}\right)\label{idef}
 \eea
 Numerical evaluation leads to the result 
 \bea I({\rm electron})=1.7,\quad  I({\rm muon})=0.17,\label{res} \eea
 and the virtual $e^+e^-$ pair  contribution to the magnetic moment of the quark is given by 3.4 $({\alpha\over \pi})^2$ or $1.8\times 10^{-5}$.
 This number is about 100  times larger than the value of $\epsilon_p$ used to account for the proton radius puzzle.  Moreover, if the value of the lepton mass in the loop were to dominate  the value we would expect a muon to electron ratio of about 1/200 instead of the ratio $\sim 1/10$ of the values of \eq{res}. 
 
It is useful to attempt  to relate the results of \eq{res} to access the value of $\epsilon_p$, even though this can only be done heuristically. In non-relativistic
 quantum mechanics the contribution of a component $n$ to the 
expectation value of  an operator  $ {\cal O} $ is written as $\langle {\cal O} \rangle=\int d^3r \psi_n^*(\bfr) {\cal O} \psi_n(\bfr)$. This  can be interpreted as $\int d^3r|\psi_n(\bfr)|^2\times \bar{\cal O}= P_n\bar{\cal O}$.
 where $\bar{\cal O}$  is the average of $\cal O$ in the component $n$. If this operator is written in natural dimensionless units, such as  a coefficient multiplying the factor $\bar{u}(p'){i\sigma^{\mu\nu}q_\nu\over 2m}u(p)$, then $\bar{\cal O}$ can be expected to be of the order of unity. In that case the numerical coefficient, here $1.8\times 10^{-5}$ can be expected to be of the order of the probability for  the electron-postiron pair to exist. Thus, we state, very roughly, that 
 \bea \epsilon_p\sim1.8\times 10^{-5}.\label{mag}\eea 

 Note that the dimensionless  factor, $I$, multiplying $\bar{u}(p'){i\sigma^{\mu\nu}q_\nu\over 2m}u(p)$  depends on the ratio of the quark to lepton masses. This is not surprising-at $q^2=0$ the only parameters with dimension are $m,m_l$ and the only way to make a dimensionless number is a dependence on the ratio.
 The consequences of this are  discussed below in Sect.~\ref{ass}.
\section{Axial  coupling}

We seek  another example of a divergent-free  matrix element that depends on the lepton-pair content of the proton.  Consider the axial coupling $\gamma^\mu\gamma^5$ operator as an insertion in Fig.2b.
The resulting term is defined as $\delta A^\mu$, with 
\bea 
\delta A^\mu=ie^2\int {d^4k\over (2\pi)^4}{\bar{u}(p')\gamma_\rho\left (\slashed{k}'+m)\gamma^\mu\gamma^5(\slashed{k}+m)  \right)\gamma^\rho u(p)\over \left((p-k)^2+i\epsilon)({k'}^2-m^2+i\epsilon)(k^2-m^2+i\epsilon)\right)}\hat{{\Pi}}_2((p-k)^2),\label{da}
\eea
Using    parity conservation and time reversal invariance tells us that $\delta A^\mu$ takes the form
\bea 
\delta A^\mu=\bar{u}(p')\left[G_A(q^2)\gamma^\mu\gamma^5+i {G_T(q^2)\over 2m}\sigma^{\mu\nu}q_\nu\gamma^5+{G_P(q^2)\over 2m} q^\mu\gamma^5\right]u(p).
\eea
The term $G_A$, as computed from \eq{da} will contain  an ultraviolet divergence, and are determined after a renormalization procedure. The terms $G_T,\,G_A$ are free of such  problems, so we will examine only those terms.  We again  take  $q^2=0$.

We proceed by using \eq{good} in \eq{da} to find
\bea 
\delta A^\mu= ie^2{2\alpha\over\pi}\int_0^1dz\,z(1-z)\int_0^1{d\lambda\over \lambda}\int {d^4k\over (2\pi)^4}{\bar{u}(p')\gamma_\rho\left (\slashed{k}'+m)\gamma^\mu\gamma^5(\slashed{k}+m)  \right)\gamma^\rho u(p)\over \left((p-k)^2-\hat{m}_l^2+i\epsilon)({k'}^2-m^2+i\epsilon)(k^2-m^2+i\epsilon)\right)} .\label{da1}
\eea

We proceed by combining the propagators, shifting the origin of integration, integrating over the shifted four-momentum,  integrating over one of the Feynman parameters, taking $q^2=0$ and keeping only the contribution to $G_T,G_P$. The result is 
\bea& 
\delta A^\mu_{T,P}(q^2=0)= {1\over2}({\alpha\over\pi})^2\,m\int_0^1dz\,z(1-z)\int_0^1{d\lambda\over \lambda}\int_0^1dz' \int_0^{1-z'} dy 
{\left(q^\mu \gamma^5(2+z'(1+z'))+i\sigma^{\mu\nu}q_\nu\gamma^5(-2yz'+z'(1-z')\right)
\over (1-z')^2 m^2+z' \hat{m}_l^2+i\epsilon} .\nonumber\\&
\label{da2}
\eea
The integration over $y$ gives a factor of $1-z'$ to to term proportional to $q^\mu\gamma^5$, but leads to a cancellation in the term proportional to $i\sigma^{\mu\nu}q_\nu\gamma^5$. Thus the $G_T$ term vanishes at $q^2=0$. This cancellation does not occur for other values of $q^2$. The integration over $\lambda$ is performed to give the result:
\bea
&\delta A^\mu_{P}(q^2=0)= {1\over2}\left({\alpha\over\pi}\right)^2\,{q^\mu \gamma^5\over m}J(m^2/m_l^2)\\
&J(m^2/m_l^2)\equiv\int_0^1dz\,z(1-z)\int_0^1dz'  
{((2+z'(1+z'))\over 1-z'}\log[1+{(1-z')^2z(1-z)\over z'}{m^2\over m_l^2}]\label{jdef}
\eea
The remaining integrals are handled numerically.
The result is that   
\bea J(q^2=0,{\rm electron})=18.3,\quad  J(q^2=0,{\rm muon})=0.593,\label{res} \eea
We again see that the effect of $e^+e^-$  pairs is much larger than  the 2 $\times 10^{-7}$ that is supposed to enter in the proton radius puzzle.  Using the logic of the previous section leads to the approximate relation 
\bea \epsilon_p\sim 5\times 10^{-5}.\label{ax}\eea 
This has about 250 times larger than the value used in Ref.~\cite{Jentschura:2014ila}.
The numbers displayed in \eq{mag} and \eq{ax} are of the same order of magnitude, which is  all that can be expected from the quantitative approach used here.
\section{Assessment}
\label{ass}
We have provided arguments that  the lepton pair content is an order of magnitude  larger than the value of $\epsilon_p$ needed to account for the proton radius puzzle. 
One might expect that  this strengthens the case for the explanation of Ref.~\cite{Jentschura:2014ila}. 

\begin{figure}[h]
\includegraphics[width=7cm,height=6cm]{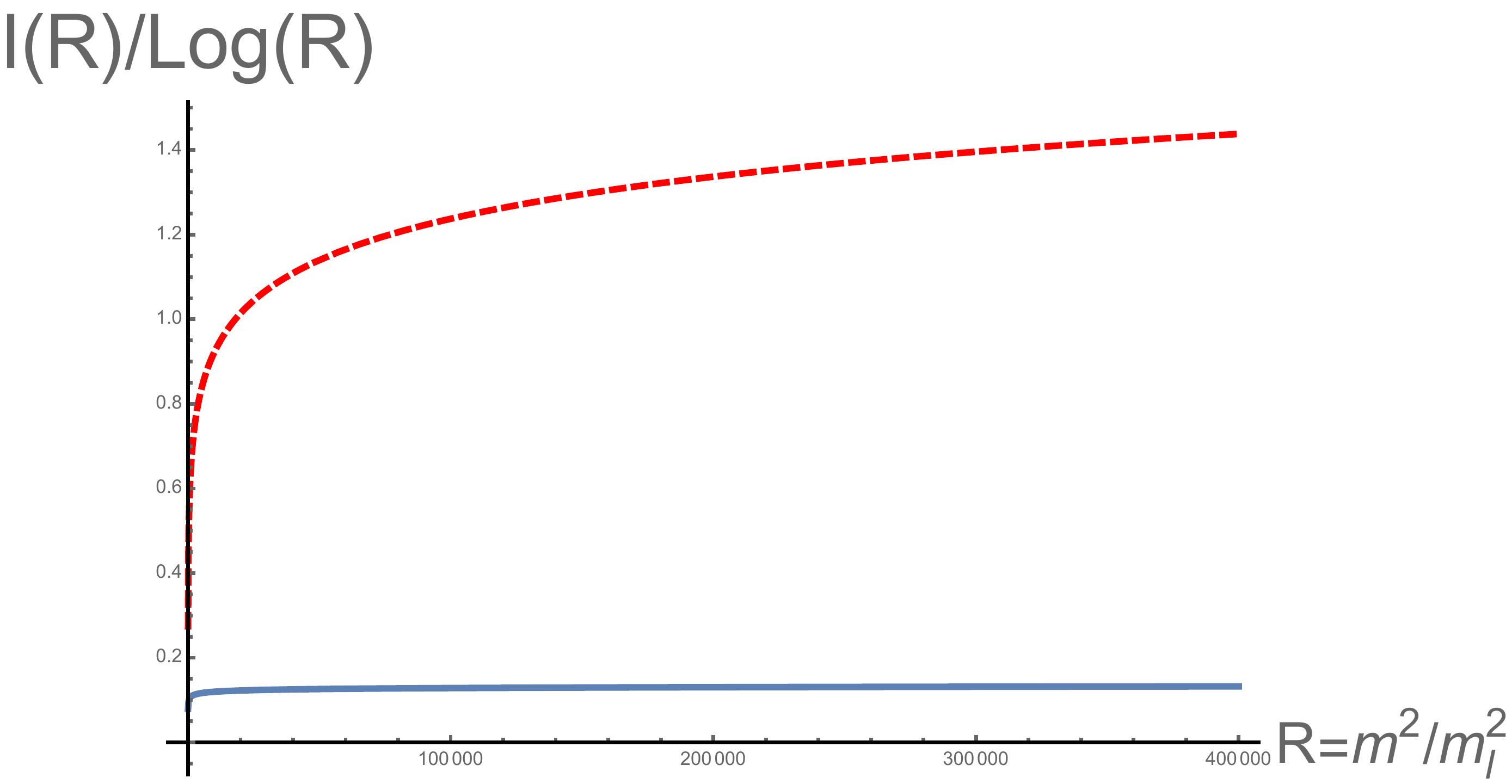}
\caption{(color online) Dependence on the ratio $m^2/m_l^2$.  The functions $I(m^2/m_l^2)/\log(m^2/m_l^2)$,\eq{idef} (blue, solid) and $J(m^2/m_l^2)/\log(m^2/m_l^2)$, \eq{jdef} (red, dashed) are displayed.
 }
\label{log}\end{figure}

That this is not so can be understood by examining the validity of \eq{ann},  derived by assuming that the electron and positron annihilate when both are at rest. Firstly, if this equation is correct, and the lepton-pair  probabilities are as large as obtained in this paper, then the computed effect would be between 100 and 250 times times too big., causing a disagreement with experiment that  would rule out the explanation of  Ref.~\cite{Jentschura:2014ila}. 
But it is necessary to  assess the lepton  mass dependence that arises from evaluating the loop diagrams. This is in both examples approximately a logarithm of the ratio $m^2/m_l^2$, as shown in Fig.~\ref{log}. This means that a dependence as the inverse square of the lepton mass will not result from evaluating any loop diagram.

Moreover, we may use dimensional analysis to  argue that a complete evaluation of the effect depicted in Fig.~1 would lead to a $1/m^2$ dependence. This is because the
inverse factor of  $4m_e^2$ appearing in \eq{one} arises from  the propagator of   the  virtual photon  that is produced by the annihilation. In the diagram this factor would be 
$(p_e^-+p_e^+)^2$. The momentum $p_e^-$ of the external electron  is given almost entirely by the electron mass, but the four-momentum of the space-like  virtual positron,  $p_e^+$, is dominated by its three-momentum. This in turn is governed in loop integrals  by the largest mass, which is that of the quark, $m$. In particular, 
a correct and complete evaluation would lead one to obtain
\begin{equation}
\delta H_{\rm ann} \propto \frac{\pi \alpha}{2 m^2} \, 
\left( 3 +  \boldsigma_+ \cdot \boldsigma_-
\right)  \, \delta(\bfr) \,
\end{equation}
instead of \eq{one}.
Since $m\,\approx  600 \,m_e$  the effect  is about four hundred thousand times smaller than what is needed and therefore can be said to vanish.

We have used the term ``a full and complete evaluation". What would be needed for that? The minimum requirement is  gauge invariance.  That the  diagram of  Fig. 1  can not satisfy gauge invariance by itself is apparent because there are many other diagrams of the same  order.
In particular, there is a diagram corresponding to crossing of the photons emitted by the $u$ and $d$ quarks of  Fig~1.  Terms with both  crossed and uncrossed photons are needed to satisfy gauge invariance in Compton scattering. This means that the 
lepton pair content can not  be evaluated  without also including the effects of the $2\gamma\,e^-\,e^+$ component. 
One simply needs to compute the complete gauge invariant set of diagrams corresponding to those implied in Fig~2. This has been done for the case of muonic hydrogen. Given the insensitivity of the results to the value of the lepton mass, we expect that  for electronic hydrogen the lepton-pair effect will be of order of  $\alpha/\pi$ of the proton polarizability correction and therefore negligible.

\section{Nuclear Dependence}
\label{nuc}

Models such as the contact interactions of \eq{ann}  which behave as a delta function in the separation between the lepton and nucleon contain  very specific 
predictions for the  nuclear dependence of the Lamb shift.

The first example we consider is the model of Ref.~\cite{Jentschura:2014ila}.  To make a prediction for the nucleus, one needs
 to know the contribution of the neutrons. Using the single-quark dominance idea of Sect~II  gives a specific result obtained by considering the factors of the square of the  quark charge that would
 appear in the calcuation.
A  proton  contains  two up quarks and one down quark, with a resulting
 quark charge squared factor of  $2 (2/3)^2e^2+1(1/3)^2e^2=e^2$ . For a neutron one would have  $1 (2/3)^2e^2+2(1/3)^2e^2=2/3e^2$.
Thus the neutron contribution would be 2/3 that of the proton.
As result, if one considers the Lamb shift in the electron-deuteron atom, the effect would  5/3 as large as for a proton. Such an effect would contradict the existing good agreement between theory and experiment~\cite{Pohl:2011pxa}.

More generally, suppose the contribution of a proton to the Lamb shift is $E_p$ (0.3meV)  to resolve the proton radius puzzle) and that of the neutron is $E_n$.  
Then for a nucleus with $A$ nucleons and $Z$ protons,  we find 
\bea
E_A=\left({1+{m_\mu\over m_p}\over 1+{m_\mu\over Am_p}}\right)^3Z^3 (ZE_p+N E_n)\left(1-{\cal O}(\,{R_A^2\over a_\mu^2}\,)\right)\approx \left({1+{m_\mu\over m_p}\over 1+{m_\mu\over Am_p}}\right)^3Z^3 (ZE_p+N E_n),\label{NUC}\eea
where $a_\mu$ is the muon Bohr radius ($>$100 times larger than nuclear radius, $R_A$).  The meaning of \eq{NUC} is  that the contributions of  such contact interactions increase very rapidly with atomic number.

In particular,  the prediction of Ref.~\cite{Jentschura:2014ila}(with $E_n=2/3E_p$) for $^4$He is a Lamb shift 
that is (1.27) 8 (2)5/3$\approx$   10 meV, a huge number.
The expression \eq{NUC} applies to all models in which the contribution to the Lamb shift enters as a delta function (or of very short range) in the lepton-nucleon coordinate, including ~\cite{Miller:2012ne,Wang:2013fma}.
 In Ref.~\cite{Miller:2012ne}, which concerns polarizability corrections, the neutron contribution, $E_n$  could vanish, so that the contribution for $^4He$ would be
 20 $E_p$=6 meV.
 In Ref.~\cite{Wang:2013fma}, which concerns a gravitational effect, $E_n=E_p$, so the prediction for $^4He$ would be
 40$E_p$=12 meV. These various predictions will be tested in an upcoming experiment~\cite{Nebel:2012qla}.
It is also worth mentioning the MUon proton Scattering Experiment (MUSE)~\cite{Gilman:2013eiv},  a simultaneous
measurement of $\mu^\+\,p$ and $e^+\,p$  scattering and also a simultaneous
measurement of $\mu^\-\,p$ and $e^-\,p$  scattering that is particularly sensitive to the presence of contact interactions~\cite{Miller:2012ne} .

 \section{Summary}
The work presented here supports the idea of the existence of a non-perturbative lepton-pair content of the proton. Such components are not forbidden by any symmetry principle 
and therefore must appear. However, the presented calculations show that such a content  is not  a   candidate to  explain  the proton radius puzzle. This is because 
computation of the necessary loop effects is cannot yield an effective Hamiltonian of the  strength and form of \eq{ann}. The $1/m_l^2$ behavior of that equation  in Ref.~\cite{Jentschura:2014ila} is necessary to obtain the needed magnitude of the separate electron and muon  Lamb shifts. Instead, loop calculations are expected to lead to a  a dependence of $1/m^2$, with $m$ the constituent quark mass.  This means that the effect of  Ref.~\cite{Jentschura:2014ila} is expected to be entirely negligible. This is in accord with the estimate: ($\alpha/\pi$) times the polarizability correction that is obtained from evaluating the relevant Feynman diagrams.
 
 More generally: it can be said that the proton does have  Fock -space components
containing lepton pairs. However, the simplest and most  reliable method of treating such  pairs is to compute gauge-invariant sets of Feynman diagrams using QED perturbation theory.  
\section{Acknowledgements}
  This material is based upon work supported by the U.S. Department of Energy Office of Science, Office of Basic Energy Sciences program under Award Number DE-FG02-97ER-41014.
I thank S. Paul, U. Jentschura, S. Karshenboim and M. Eides  for useful discussions.

\end{document}